\documentclass[aps,pra,superscriptaddress,onecolumn]{revtex4}

\usepackage{amsfonts}
\usepackage{amssymb}
\usepackage{graphicx}
\usepackage{bbm}
\usepackage{amssymb}
\usepackage{graphicx}
\usepackage{amsmath}
\usepackage{amsbsy}
\usepackage{amsthm}
\usepackage{epsfig}

\begin{document}

\title{Can a spontaneous collapse in flavour oscillations be tested at KLOE?}

\author{Kyrylo Simonov}
\affiliation{University of Vienna, Faculty of Physics, Boltzmanngasse 5, 1090 Vienna, Austria.}
%\email{Kyrylo.Simonov@univie.ac.at}
\author{Beatrix C. Hiesmayr}
\affiliation{University of Vienna, Faculty of Physics, Boltzmanngasse 5, 1090 Vienna, Austria.}
\email{Beatrix.Hiesmayr@univie.ac.at}

\begin{abstract}Why do we never see a table in a superposition of here and there? This problem gets a solution by so called collapse models assuming the collapse as a genuinely physical process. Here we consider two specific collapse models and apply them to systems at high energies, i.e. flavour oscillating neutral meson systems. We find on one hand a potentially new interpretation of the decay rates introduced by hand in the standard formalism and on the other hand that these systems at high energies constrain by experimental data the possible collapse scenarios.
\end{abstract}

\maketitle

\section{Introduction: The Measurement Problem}
Quantum mechanics is an exceedingly successful theory which can explain a plethora of experimental results covering physical phenomena on different energy scales. However, in its standard formulation quantum mechanics  meets some conceptual problems. For instance, formally a macroscopic object can be in a superposition as the famous example of the Schrödinger's cat. The interaction of a seemingly macroscopic system such as a measurement apparatus with a seemingly microscopic system, a quantum system, is postulated to force a reduction/collapse of the wave function of the quantum system, i.e. breaking up the superposition. Standard quantum theory neither provides a mechanism how this collapse takes place nor reveals whether it is a real physical process. We have two completely different types of dynamics at hand, i.e.
\begin{itemize}
 \item a deterministic unitary time evolution ruled by the Schr\"{o}dinger equation and
 \item a stochastic and non-unitary reduction process of the wave function caused by a measurement obeying Born's rule.
\end{itemize}
Obviously, several questions arise: What defines a system to be microscopic versus macroscopic? The Copenhagen interpretation uses this concept but never defines it. Is the collapse procedure a real physical process? Does the collapse manifest at high energies differently than for systems at lower energies? Do collapse models allow for a different interpretation of the measurable dynamics of flavour oscillating mesons? In turn, how do these systems at high energies restrict the plethora of collapse models?

Dynamical reduction models (or collapse models) claim to solve this issue by modeling the collapse as a genuinely physical process. The aim of this contribution is to investigate collapse models for oscillation systems at high energies such as the K-mesons produced by $\mbox{DA}\Phi\mbox{NE}$. The paper is organised by giving a short introduction to collapse models and then applying two popular models to neutral meson systems followed by interpreting the results.

\section{Collapse models: A Solution to the Measurement Problem?}
Dynamical reduction models provide one of the possible solutions to the measurement problem. In general they propose a new universal dynamics which introduces the collapse of the wave function as an objective physical process. In order to recover the predictions of quantum mechanics this dynamics should fulfill the following properties
\begin{itemize}
 \item non-linearity: The new dynamics should break superpositions on a macroscopic level, particularly, during a measurement.
 \item stochasticity: The new dynamics should produce the quantum probabilities obeying Born's rule.
 \item no superluminal signaling: The new dynamics should not be in conflict with the special relativity.
 %\item experiment: The new dynamics should be in no conflict with all up to now performed experiments.
\end{itemize}
Let us mention that it is by no means trivial that a mathematically consistent framework exists with the above properties and simultaneously agrees with all observed data. The first dynamical reduction model was the GRW model introduced in 1985 by Ghirardi, Rimini and Weber~\cite{GRW}. It proposes a simple mechanism for the collapse for a system of $N$ particles through spontaneous localisations, i.e.
\begin{enumerate}
 \item Each particle undergoes a sudden localisation at a random time $t$ (mathematically  modelled by a multiplication with a peaked Gaussian function).
 \item Between such localisations the state of the system evolves due to the Schr\"{o}dinger equation.
\end{enumerate}
In this way the GRW model, and collapse models in general, introduce two new fundamental constants, the localisation rate $\lambda_{GRW}$, which fixes the mean rate of the localisation process, and the coherence length $r_C$, which fixes the width of the Gaussian function, i.e. the localisation effect. These two constants allow to divide the world into a microscopic regime, where the Schr\"{o}dinger equation rules the dynamics mainly, and a macroscopic regime where superposition breaks (Schr\"{o}dinger's cat is alive). Consequently, taking collapse models seriously there are two new natural constants ruling our world.

Around 1989 two more realistic dynamical reduction models appeared on the market. These are the QMUPL model (Quantum Mechanics with Universal Position Localisation)~\cite{Diosi1989} and the CSL model (Continuous Spontaneous Localisation)~\cite{Pearle1989, Ghirardi1990, GhirardiGrassiBenatti1995}. In contrast to the GRW model these models introduce the spontaneous collapse as a continuous process. The dynamics is obtained by adding to the Schr\"{o}dinger equation non-linear and stochastic terms. For a given Hamiltonian $\hat{H}$ and state $|\psi_{t}\rangle$ at time $t$ the dynamics is defined by~\cite{BassiDuerrHinrichs2011}
\begin{eqnarray}\label{StateVectorEquation}
d|\psi_{t}\rangle&=&\Bigl[-i\hat{H}\, dt+\sqrt{\lambda}\sum\limits_{i=1}^N(\hat{A}_i - \langle \hat{A}_i \rangle_t) dW_{i, t} -\frac{\lambda}{2} \sum\limits_{i=1}^N(\hat{A}_i- \langle \hat{A}_i \rangle_t)^2dt\Bigr]\;|\psi_{t}\rangle,
\end{eqnarray}
where $\hbar=1$ is taken, $\langle \hat{A}_i \rangle_t := \langle \psi_t | \hat{A}_i |\psi_{t}\rangle$ is the quantum mechanical expectation value, and $\lambda$ sets the strength of the collapse (connected to the rate and coherence length). Here $\hat{A}_i$ represent a set of $N$ self-adjoint commuting operators introducing the collapse, $W_{i,t}$ represent a set of $N$ independent standard Wiener processes, one for each collapse operator $\hat{A}_i$, and lead to a white noise field $w_{i,t}=\frac{d}{dt} W_{i,t}$ which plays a crucial role in the model. The two collapse models differ mainly in the choice of the collapse operators $\hat{A}_i$.

Both models have maintained a high interest over the last years being intensively investigated from the fundamental point of view as well as in experiments. Particularly, the QMUPL model has been investigated by analysing the spontaneous radiation emission from a free charged particle~\cite{BassiDuerr2009, BassiDonadi2014} and X-ray emission by an isolated slab of Germanium~\cite{CurceanuHiesmayrPiscicchia2015, Curceanu2015, Piscicchia2015}. For the CSL model experiments with optomechanical cavities~\cite{BahramiPaternostro2014, NimmrichterHornberger2014} and mechanical oscillators~\cite{Diosi2015} were proposed, contributions from cold-atom experiments~\cite{Billardello2016} were also recently obtained.

In the high energy regime the first computations for neutral meson and neutrino systems can be found in Refs.~\cite{Donadi2013, Bahrami2013}. These predictions were compared to specific decoherence models~\cite{Deco1Bertlmann1999, Deco2Hiesmayr2001, Deco3Bertlmann2003}, where the decoherence strength was computed by KLOE data~\cite{Ambrosino2006, DiDomenico2008, DiDomenico2010} and  by data from B-factories~\cite{Go2007, Richter2008, Yabsley2008}. Last but not least it was shown that the collapse models can contribute to an effective cosmological constant~\cite{Josset2016}. Thus the dynamical reduction models reveal a rich spectrum of potential experimental tests.

\section{Neutral Mesons: A Laboratory for Testing the Collapse Models}

\begin{figure}
\minipage{0.48\textwidth}
\includegraphics[width=0.9\textwidth]{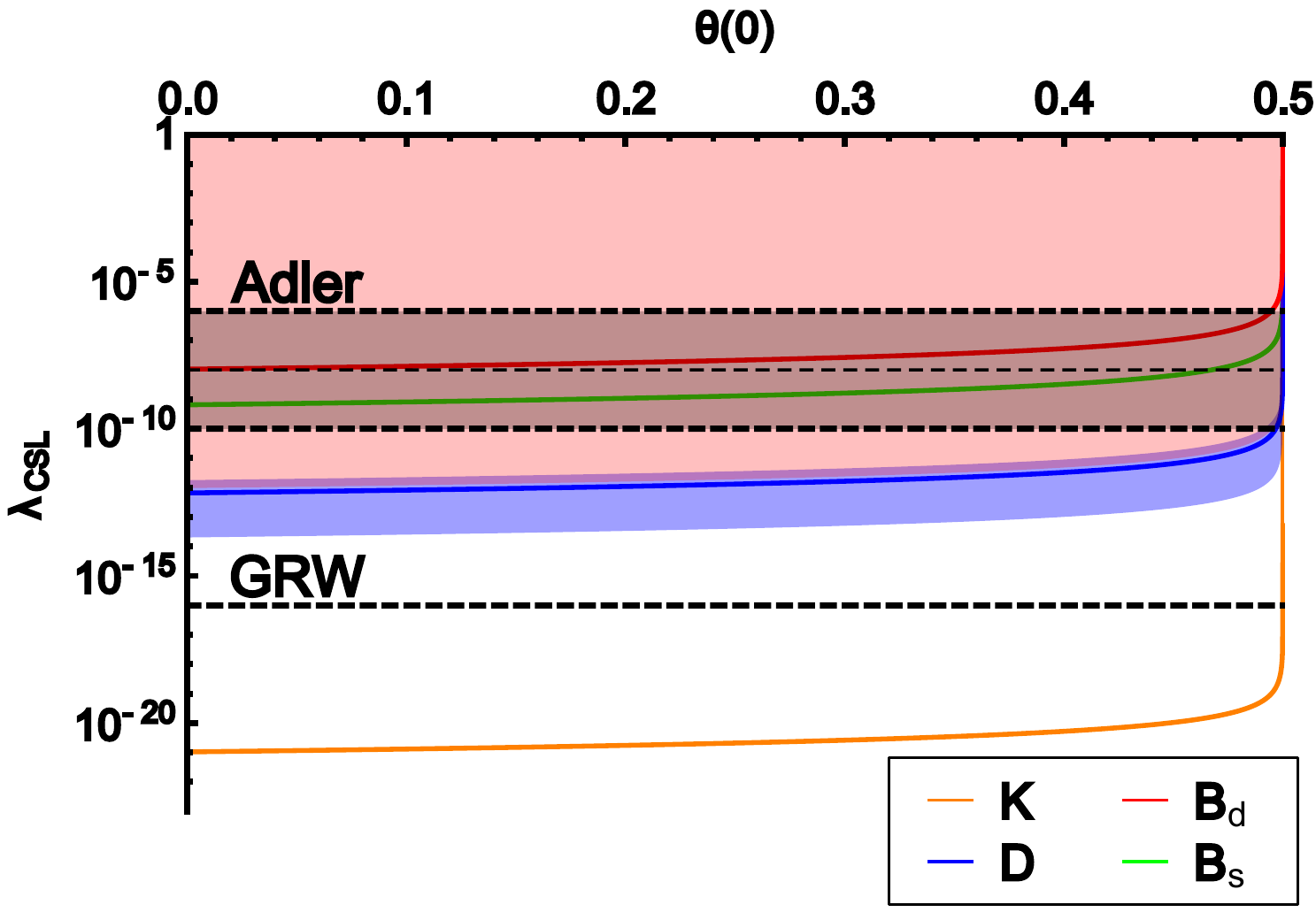}
\caption{Values of the collapse rate including experimental errors as a function of $\theta(0)$ based on the values $\Gamma_H^{\textrm{exp}},\Gamma_L^{\textrm{exp}},\Delta m^{\textrm{exp}}$ known from experiments for the different types of neutral mesons. The plot is taken from~\cite{SimonovHiesmayrPaper2016}.}\label{ThetaVSRate}
\endminipage\hfill
\minipage{0.48\textwidth}%
\includegraphics[width=0.8\textwidth]{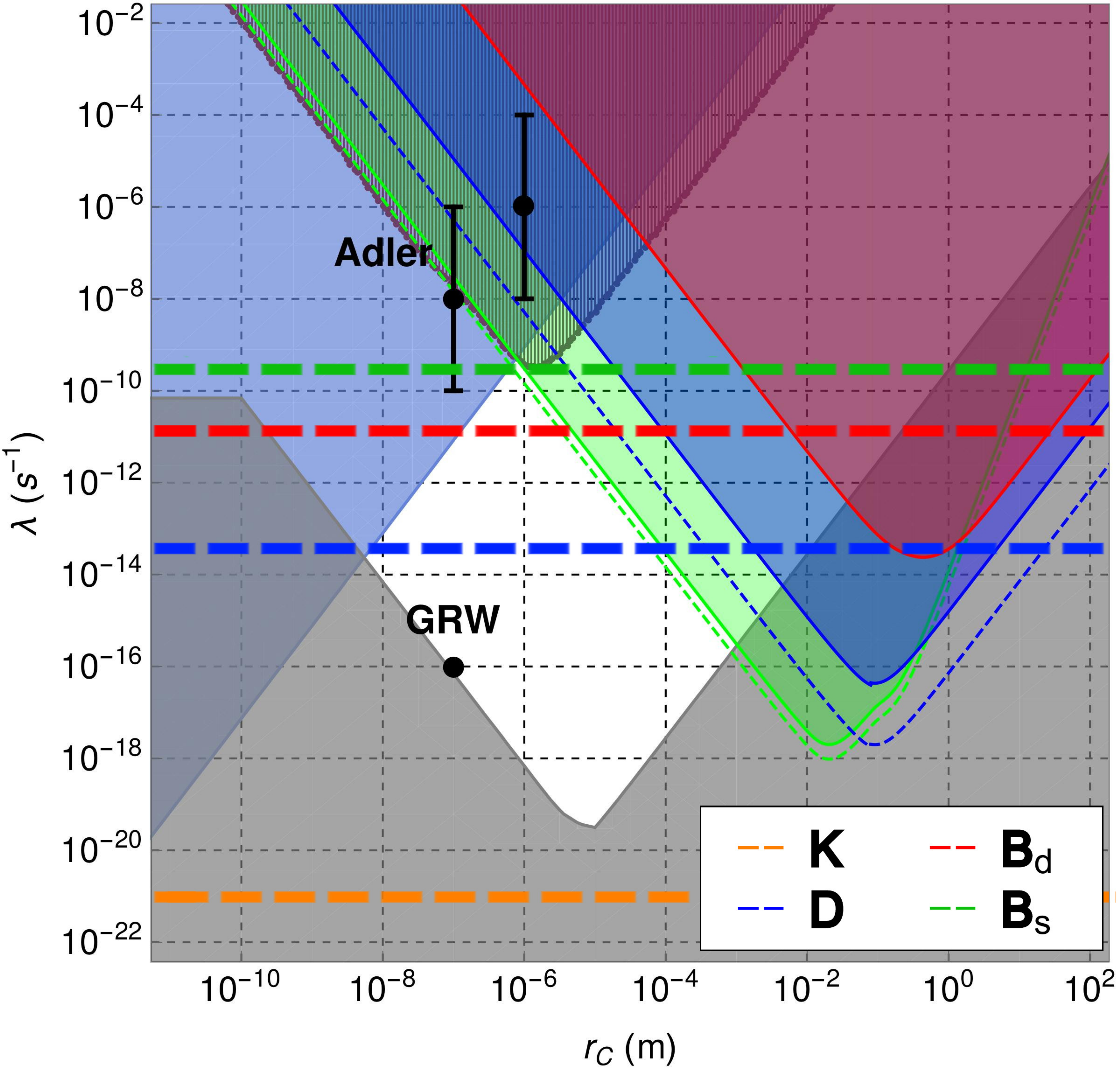}
\caption{Bounds on the natural constants of the CSL model based on LIGO, LISA Pathfinder and AURIGA experiments (blue, green, and red lines), ultracold cantilever experiments (purple line), X-ray experiments (light blue line) and theoretical estimations (grey line) (taken from Ref.~\cite{CarlessoBassi2017}). The thick dashed lines refer to the lower bounds due to our computations of $\lambda^{CSL}$, Eq.~(\ref{resultlambda}), for the neutral meson system. Note that $r_C$ is not bounded.}\label{BassiRates}
\endminipage
\end{figure}

Flavour oscillating systems, in particular the neutral $K$-mesons, have proven to be an exceptional laboratory for exploring quantum foundations, for instance a puzzling connection between the violation of the $\mathcal{CP}$ symmetry and the violation of a Bell inequality~\cite{BellKaonHiesmayr2012} has been found (experimentally feasible with the KLOE detector). Neutral mesons are massive systems with two different mass eigenstates $|M_L\rangle$ and $|M_H\rangle$ which decay with distinct decay rates. The difference between decay rates for $K$-mesons is rather huge, in strong contrast to the other mesons. In what follows we neglect the tiny violation of the $\mathcal{CP}$ symmetry (which means that $\langle M_H|M_L\rangle= 0$). A neutral meson consists of quark-antiquark pair, and both the particle state $|M^0\rangle$ and the antiparticle state $|\bar M^0\rangle$ can decay into the same final states. Therefore, neutral mesons have to be considered as a two-state system and their dynamics is then described by the following effective Schr\"{o}dinger equation due to the Wigner--Weisskopf approximation
\begin{align}\label{WignerWeisskopf}
 & \frac{d}{dt}\; |\psi_t\rangle = - i\, \hat{H}\; |\psi_t\rangle \implies |\psi_t\rangle = a(t)\; |M^0\rangle + b(t)\; |\bar{M}^0\rangle,
\end{align}
where the effective Hamiltonian $\hat{H} = \hat{M} + \frac{i}{2} \hat{\Gamma}$ is non-hermitian, and $\hat{M}$ is the hermitian mass operator, and $\hat{\Gamma}$ is a hermitian operator which describes the decay. The diagonalised Hamiltonian defines the mass eigenstates
\begin{align}\label{MassStates}
 & \hat{H}\; |M_n\rangle = \Bigl(m_n + \frac{i}{2} \Gamma_n \Bigr)\; |M_n\rangle,
\end{align}
where $\Gamma_L, \Gamma_H$ are the corresponding decay rates (with $c=1$). The flavour eigenstates are related to the mass eigenstates by
\begin{subequations}
 \begin{eqnarray}\label{FlavourStates}
  & |M^0\rangle = \frac{1}{\sqrt{2}} \Bigl( |M_H\rangle + |M_L\rangle \Bigr), \\
  & |\bar{M}^0\rangle = \frac{1}{\sqrt{2}} \Bigl( |M_H\rangle - |M_L\rangle \Bigr).
 \end{eqnarray}
\end{subequations}
The general collapse scenario, Eq.~(\ref{StateVectorEquation}), has been shown to be in our case equivalent to~\cite{AdlerBassi2007, SandroPhD}
\begin{equation}\label{ImTrickStateVectorEquation}
i \frac{d}{dt} |\psi_{t}\rangle = \Bigl[ \hat{H} - \sqrt{\lambda} \sum\limits_{i=1}^N \hat{A}_i w_{i, t} \Bigr]\; |\psi_{t}\rangle := \Bigl[ \hat{H} + \hat{N}(t) \Bigr] \; |\psi_{t}\rangle,
\end{equation}
which has a form of a standard Schr\"{o}dinger equation with a random perturbation $\hat{N}(t)$.
Solving this equation~(\ref{ImTrickStateVectorEquation}) with the help of Dyson series up to second order we obtained after a cumbersome computation the following probabilities~\cite{SimonovHiesmayrLetter2016, SimonovHiesmayrPaper2016} for the propagation of the mass eigenstates for the QMUPL and CSL models
\begin{align}
& P^{QMUPL}_{M_{\mu=L/H}\rightarrow M_{\nu=L/H}} (t) = \delta_{\mu\nu}\;\Bigl( 1 -  \Lambda^{QMUPL}_\mu\cdot t + 3\cdot\frac{1}{2}(\Lambda^{QMUPL}_\mu)^2\cdot t^2 \Bigr)\cdot e^{-\Gamma_\mu t}, \\
& P^{CSL}_{M_{\mu=L/H}\rightarrow M_{\nu=L/H}} (t) = \delta_{\mu\nu}\;\Bigl( 1 -  \Gamma^{CSL}_\mu\cdot  t + \frac{1}{2}(\Gamma^{CSL}_\mu)^2 \cdot t^2 \Bigr)\cdot e^{-\Gamma_\mu t} \approx \delta_{\mu\nu}\; e^{-(\Gamma^{CSL}_\mu+\Gamma_\mu) t},
\end{align}
and for the propagation of the flavour states
\begin{eqnarray}
P^{QMUPL}_{M^0 \rightarrow M^0/\bar{M}^0} &=& \frac{1}{4} \Biggl\{ \sum_{i=H,L} e^{-\Gamma_i t}\; \Bigl( 1 -  \Lambda^{QMUPL}_i\cdot t + 3\cdot\frac{1}{2}(\Lambda^{QMUPL}_i)^2\cdot t^2 \Bigr)\nonumber\\
 && \pm 2\cos(\Delta m t)\;e^{-\frac{\Gamma_H+\Gamma_L}{2} t} \cdot \left(1-\frac{\alpha \lambda}{2} \left[\frac{\Delta m^2}{m_0^2} (1-\theta(0)) + \frac{m_H m_L}{m_0^2} (1-2 \theta(0))\right]\cdot t\right.\nonumber\\
 &&\left.\qquad + \; 3\cdot\frac{1}{2}\left(\frac{\alpha \lambda}{2} \left[\frac{\Delta m^2}{m_0^2} (1-\theta(0))+ \frac{m_H m_L}{m_0^2} (1-2 \theta(0))\right]\right)^2\cdot t^2\right)\Biggr\},\\
 P^{CSL}_{M^0 \rightarrow M^0/\bar{M}^0} &=& \frac{1}{4} \Biggl\{ \sum_{i=H,L} e^{-\Gamma_i t}\; \Bigl( 1 -  \Gamma^{CSL}_i\cdot t + \frac{1}{2}(\Gamma^{CSL}_i)^2\cdot t^2 \Bigr)\nonumber\\
 && \pm 2\cos(\Delta m t)\;e^{-\frac{\Gamma_H+\Gamma_L}{2} t}\cdot \left(1-\frac{\gamma}{(\sqrt{4\pi}r_C)^d} \left[\frac{\Delta m^2}{m_0^2} (1-\theta(0))+\frac{m_H m_L}{m_0^2} (1-2 \theta(0))\right]\cdot t\right.\nonumber\\
 \nonumber &&\left.\qquad +\frac{1}{2}\left(\frac{\gamma}{(\sqrt{4\pi}r_C)^d} \left[\frac{\Delta m^2}{m_0^2} (1-\theta(0))+\frac{m_H m_L}{m_0^2} (1-2 \theta(0))\right]\right)^2\cdot t^2\right)\Biggr\} \\
 &\approx& \frac{e^{-(\Gamma_L+\Gamma^{CSL}_L)t}+e^{-(\Gamma_H+\Gamma^{CSL}_H)t}}{4}\cdot \Biggl\{ 1\pm \frac{\cos(\Delta m t)}{\cosh(\frac{(\Gamma_L+\Gamma^{CSL}_L)-(\Gamma_H+\Gamma^{CSL}_H)}{2}\cdot t)}\cdot e^{- \frac{\gamma}{ (\sqrt{4\pi} r_C)^d}\frac{(\Delta m)^2}{2\,m_0^2} t} \Biggr\}\;,\nonumber\\
\end{eqnarray}
where we used the abbreviations
\begin{eqnarray}
\Lambda^{QMUPL}_\mu = \frac{\alpha\lambda}{2}\cdot \frac{m_{\mu}^2}{m_0^2}\cdot  \Bigl( 1 - 2\theta(0)\Bigr)\end{eqnarray} and
\begin{eqnarray}\Gamma^{CSL}_\mu= \frac{\gamma}{(\sqrt{4\pi}r_C)^d}\cdot \frac{m_{\mu}^2}{m_0^2}\cdot \Bigl(1 - 2\theta(0)\Bigr)\;.\end{eqnarray}

Here $d$ is the dimensionality, $\lambda = \frac{\lambda_{GRW}}{2r_C^2}$ is the collapse strength in the QMUPL model, $\gamma = \lambda_{GRW} \cdot (\sqrt{4\pi}r_C)^d$ is the collapse strength in the CSL model, $m_0$ is a reference mass which is taken usually to be the nucleon mass (we will use the rest mass of the corresponding neutral meson), $\Delta m = m_H - m_L$ is the mass difference, $\theta(0)$ is the Heaviside function at zero, and $\sqrt{\alpha}$ represents a width of the wave packet that is assumed to be the initial state of the meson for the QMUPL model.

The obtained probabilities reveal several interesting results, namely:
\begin{itemize}
 \item For the QMUPL model the probabilities have a non-exponential contribution which makes the collapse effect in principle observable.
 \item For the CSL model the probabilities can be expanded to exponentials, allowing for new interpretations.
 \item The collapse dynamics contributes to the transition probabilities by two independent effects, a damping of the oscillation and a contribution to the decay rates.
 \item While the damping of the oscillation is proportional to the experimentally measurable squared mass difference, the effective decay constants are proportional to the squared absolute masses (which do not show up in the standard quantum mechanical framework as measurable quantities).
\end{itemize}

\begin{center}
\begin{table}
  \begin{tabular}{ ||c | c | c | c | c | c || }
    \hline
    & $\Gamma_L^{\textrm{exp}}$ [$s^{-1}$]& $\Gamma_H^{\textrm{exp}}$  [$s^{-1}$]& $\Delta m^{\textrm{exp}}$  [$\hbar s^{-1}$]& $m_L$ [$\hbar s^{-1}$]& $m_H$ [$\hbar s^{-1}$]\\ \hline
    $K$-mesons & $1.117 \cdot 10^{10} $ & $1.955 \cdot 10^{7} $ &$0.529 \cdot 10^{10}$& $2.311 \cdot 10^8$ & $5.524 \cdot 10^9$  \\
    $D$-mesons &  $2.454\cdot 10^{12}$ & $2.423 \cdot 10^{12}$ &$0.950 \cdot 10^{10}$&  $1.468 \cdot 10^{12}$ & $1.477 \cdot 10^{12}$  \\
    $B_d$-mesons & $6.582 \cdot 10^{11}$ & $6.576 \cdot 10^{11} $ &$0.510 \cdot 10^{12}$& $1.020 \cdot 10^{15} $ & $1.020 \cdot 10^{15} $  \\
    $B_s$-mesons & $7.072\cdot 10^{11} $ & $6.158 \cdot 10^{11} $ &$ 1.776 \cdot 10^{13}$& $2.477 \cdot 10^{14} $ & $2.655 \cdot 10^{14} $ \\
    \hline
  \end{tabular}\caption{Experimental values of the decay rates, the mass difference and the computed values of the absolute masses for $K$-, $D$-, $B_d$- and $B_s$-mesons. The table is taken from~\cite{SimonovHiesmayrPaper2016}.}\label{AbsMasses}
  \end{table}
\end{center}

In general, the obtained probabilities are strongly sensitive to the very nature of the noise field which results in the freedom to choose the value of the Heaviside function at zero $\theta(0)\in[0,1]$. Furthermore, a fully new interpretation of the dynamics can be obtained~\cite{SimonovHiesmayrPaper2016}. The decay rates can be fully derived as an effect of the collapse. Note that in any standard framework handling the meson dynamics only the mass difference $m_H-m_L$ appears, never the absolute masses.  This can be achieved by reversing the mass ratio $\frac{m_\mu}{m_0}$ in the decay constants $\Gamma_\mu^{CSL}$ (which is more reasonable from a collapse model perspective) and taking $\theta(0)\in[0,1/2\}$ to save their positivity. Considering such a scenario allows to calculate the absolute masses of the mesons using the decay rates and mass difference measured in experiments via
\begin{eqnarray}
\frac{\Gamma^{CSL}_L-\Gamma^{CSL}_H}{\Gamma^{CSL}_L+\Gamma^{CSL}_H}\;=\;\frac{m_H^2-m_L^2}{m_H^2+m_L^2}\;=\;1-\frac{m_L^2}{m_L^2+m_L\Delta m+\frac{1}{2}(\Delta m)^2},
\end{eqnarray}
and the obtained absolute masses of some neutral mesons are listed in the Table~\ref{AbsMasses}. Thus the decay rates are functions of the absolute masses and do not need an additional introduction as in the standard approach by making the Hamiltonian non-hermitian or treating the decay as a decoherence effect~\cite{Deco4Bertlmann2006, BernabeuMavromatos2013}.

From the perspective of collapse models neutral mesons allow us to derive the CSL collapse rate in return
\begin{eqnarray}\label{resultlambda}
\lambda^{CSL}&:=&\Gamma_\mu^{\textrm{exp}}\cdot \frac{m_\mu^2}{m_0^2}\frac{1}{1-2\theta(0)} = \frac{1}{(\sqrt{\Gamma_L^{-1}}-\sqrt{\Gamma_H^{-1}})^2}\frac{(\Delta m)^2}{m_0^2}\frac{1}{1-2\theta(0)}.
\end{eqnarray}
The predicted values of the collapse rate are plotted in Fig.~\ref{ThetaVSRate} for the different types of mesons and can be compared in Fig.~\ref{BassiRates} with the experimental data of different physical systems (summarised recently in Ref.~\cite{CarlessoBassi2017}).

The plots show that the values for all the mesons except for $K$-mesons are of the order of the collapse rate assumed by Adler, $\lambda_{Adler}^{CSL} =  10^{-(8\pm2)}s^{-1}$~\cite{Adler2007}. $K$-mesons tend to be closer to the original value of the collapse rate proposed by Ghirardi, Rimini and Weber, $\lambda_{GRW}^{CSL} = 10^{-16}s^{-1}$~\cite{GRW}, but require a value of $\theta(0)$ close to $\frac{1}{2}$.

\section{Conclusions}

Dynamical reduction models provide a possible solution to the measurement problem of standard quantum mechanics by introducing a physical mechanism of the collapse of the wave function. We have investigated two popular collapse models and derived their respective deviations from the standard dynamics of flavour oscillations which are intensively studied at accelerator facilities. We have found that in principle such changes are observable and the dynamics can change the oscillation behaviour and under certain constraints explain the decay behaviour of neutral mesons completely. We have also illustrated how the meson system restricts the plethora of collapse models considerably.

Precision data from experiments such as KLOE will allow further to restrict or even rule out certain collapse scenarios when one extents the dynamics to two entangled pairs of mesons or/and includes $\mathcal{CP}$ violation.\\
\\
\textbf{Acknowledgements:} Both authors acknowledge gratefully the Austrian Science Fund (FWF-P26783).


\begin{thebibliography}{1}

\bibitem{GRW}
G.~C.~Ghirardi, A.~Rimini and T.~Weber, Phys. Rev. D \textbf{34}, 470 (1986).

\bibitem{Diosi1989}
L.~Di\'{o}si, Phys. Rev. A \textbf{40}, 1165 (1989).

\bibitem{Pearle1989}
P.~Pearle, Phys. Rev. A \textbf{39}, 2277 (1989).

\bibitem{Ghirardi1990}
G.~C.~Ghirardi, P.~Pearle and A.~Rimini, Phys. Rev. A \textbf{42}, 78 (1990).

\bibitem{GhirardiGrassiBenatti1995}
G.~C.~Ghirardi, R.~Grassi and F.~Benatti, Found. Phys. \textbf{25}, 5 (1995).

\bibitem{BassiDuerrHinrichs2011}
A.~Bassi, D.~D\"{u}rr and G.~Hinrichs, Phys. Rev. Lett. \textbf{111}, 210401 (2011).

\bibitem{BassiDuerr2009}
A.~Bassi and D.~D\"{u}rr, J. Phys. A \textbf{42}, 485302 (2009).

\bibitem{BassiDonadi2014}
A.~Bassi and S.~Donadi, Phys. Lett. A \textbf{378}, 761 (2014).

\bibitem{CurceanuHiesmayrPiscicchia2015}
C.~Curceanu, B.~C.~Hiesmayr and K.~Piscicchia, J. Adv. Phys. \textbf{4}, 263 (2015).

\bibitem{Curceanu2015}
C.~Curceanu et al., Found. Phys. \textbf{46}, 263 (2016).

\bibitem{Piscicchia2015}
K.~Piscicchia et al., arXiv:1501.04462 (2015).

\bibitem{BahramiPaternostro2014}
M.~Bahrami et al., Phys. Rev. Lett. \textbf{112}, 210404 (2014).

\bibitem{NimmrichterHornberger2014}
S.~Nimmrichter, K.~Hornberger and K.~Hammerer, Phys. Rev. Lett. \textbf{113}, 020405 (2014).

\bibitem{Diosi2015}
L.~Di\'{o}si, Phys. Rev. Lett. \textbf{114}, 050403 (2015).

\bibitem{Billardello2016}
M. Bilardello et al., arXiv:1605.01891 (2016).

\bibitem{Donadi2013}
S.~Donadi et al., Found. Phys. \textbf{43}, 813 (2013).

\bibitem{Bahrami2013}
M.~Bahrami et al., Sci. Rep. \textbf{3}, 1952 (2013).

\bibitem{Deco1Bertlmann1999}
R.~A.~Bertlmann, W.~Grimus, B.~C.~Hiesmayr, Phys.~Rev.~D \textbf{60}, 114032 (1999).
%Quantum mechanics, Furry's hypothesis and a measure of decoherence in the K^0 \bar{K}^0 system

\bibitem{Deco2Hiesmayr2001}
B.~C.~Hiesmayr, Found.~Phys.~Lett. \textbf{14}, 231 (2001).
%A generalised Bell-Inequality and Decoherence for the K^0\bar{K^0}-Systems

\bibitem{Deco3Bertlmann2003}
R.~A.~Bertlmann, K.~Durstberger, B.~C.~Hiesmayr, Phys.~Rev.~A \textbf{68}, 012111 (2003).
%Decoherence of entangled kaons and its connection to entanglement measures

\bibitem{Ambrosino2006}
F.~Ambrosino et al. (KLOE collaboration), Phys. Lett. B \textbf{642}, 315 (2006).

\bibitem{DiDomenico2008}
A.~Di~Domenico et al. (KLOE Collaboration), J. Phys.: Conf. Ser. \textbf{171}, 012008 (2008).

\bibitem{DiDomenico2010}
A.~Di~Domenico et al. (KLOE Collaboration), Found. Phys. \textbf{40}, 852 (2010).

\bibitem{Go2007}
A.~Go et al. (Belle Collaboration), Phys. Rev. Lett. \textbf{99}, 131802 (2007).

\bibitem{Richter2008}
G.~Richter, {\em ``Stability of Nonlocal Quantum Correlations in Neutral B-Meson Systems''} (PhD Thesis, Technische Universit\"{a}t Wien, 2008).

\bibitem{Yabsley2008}
B.~D.~Yabsley, {\em ``Quantum entanglement at the psi(3770) and Upsilon(4S)''} (Flavour Physics \& CP Violation Conference, Taipei, 2008).

\bibitem{Josset2016}
T.~Josset, A.~Perez and D.~Sudarsky, arXiv:1604.04183 (2016).

\bibitem{BellKaonHiesmayr2012}
B.~C.~Hiesmayr et al., Eur.~Phys.~J.~C \textbf{72}, 1856 (2012).
%Revealing Bell's Nonlocality for Unstable Systems in High Energy Physics

\bibitem{AdlerBassi2007}
S.~L.~Adler and A.~Bassi, J. Phys. A \textbf{40}, 15083 (2007).

\bibitem{SandroPhD}
S.~Donadi, {\em ``Electromagnetic Radiation Emission and Flavour Oscillations in Collapse Models''} (PhD Thesis, Universit\`{a} degli studi di Trieste, 2012).

\bibitem{SimonovHiesmayrLetter2016}
K.~Simonov and B.~C.~Hiesmayr, Phys. Lett. A \textbf{380}, 1253 (2016).

\bibitem{SimonovHiesmayrPaper2016}
K.~Simonov and B.~C.~Hiesmayr, Phys. Rev. A \textbf{94}, 052128 (2016).

\bibitem{CarlessoBassi2017}
M.~Carlesso, A.~Bassi, P.~Falferi and A.~Vinante, Phys. Rev. D \textbf{94}, 124306 (2016).

\bibitem{Adler2007}
S.~L.~Adler, J. Phys. A: Math. Theor. \textbf{40}, 2935 (2007).

\bibitem{Deco4Bertlmann2006}
R.~A.~Bertlmann, W.~Grimus, B.~C.~Hiesmayr, Phys.~Rev.~A \textbf{73}, 054101 (2006).
%An open--quantum--system formulation of particle decay

\bibitem{BernabeuMavromatos2013}
J.~Bernab\'{e}u, N.~E.~Mavromatos and P.~Villanueva-P\'{e}rez, Phys. Lett. B \textbf{724}, 269 (2013).
%Consistent Probabilistic Description of the Neutral Kaon System
\end{thebibliography}
\end{document}